\documentclass{llncs}
\usepackage{tabularx}
\usepackage{sidecap}
\usepackage{comment}
\usepackage{subcaption}
\usepackage{hyperref}
\usepackage{makecell} 
\usepackage{eurosym}
\usepackage{enumitem}
\usepackage{color}
\usepackage{adjustbox}
\usepackage{subcaption}
\usepackage[misc]{ifsym}
\usepackage{url}
\DeclareCaptionType{equ}[][]

\newcommand\blfootnote[1]{%
	\begingroup
	\addtocounter{footnote}{-5}%
	\renewcommand\thefootnote{\Letter}\footnote{#1}%
	\endgroup
}

\begin{document}

\title{Predicting Knowledge Gain during Web Search based on Multimedia Resource Consumption\thanks{Part of this work is financially supported by the Leibniz Association, Germany (Leibniz Competition 2018, funding line "Collaborative Excellence", project SALIENT [K68/2017]).}}
\author{
	Christian Otto\inst{1}\protect\blfootnote{These authors contributed equally to this work.}\orcidID{0000-0003-0226-3608}\and
	Ran Yu\inst{2}\textsuperscript{\Letter}\orcidID{0000-0002-1619-3164}\and
	Georg Pardi\inst{3}\orcidID{0000-0002-7276-4099} \and 
	Johannes von Hoyer\inst{3}\orcidID{0000-0003-4306-0591} \and 
	Markus Rokicki\inst{4} \and
	Anett Hoppe\inst{1}\orcidID{0000-0002-1452-9509} \and
	Peter Holtz\inst{3}\orcidID{0000-0001-7539-6992} \and 
	Yvonne Kammerer\inst{3}\orcidID{0000-0003-4341-517X} \and 
	Stefan Dietze\inst{2,4} \and
	Ralph Ewerth\inst{1,4}\orcidID{0000-0003-0918-6297}
}
\authorrunning{C. Otto et al.}
\titlerunning{}
%
\institute{TIB -- Leibniz Information Centre for Science and Technology, Hannover, Germany\\
	\email{\{christian.otto, anett.hoppe, ralph.ewerth\}@tib.eu}
	\and
	GESIS -- Leibniz Institute for the Social Sciences, Cologne, Germany \\
	\email{\{ran.yu, stefan.dietze\}@gesis.org}
	\and
	IWM -- Leibniz-Institut für Wissensmedien, Tübingen, Germany \\
	\email{\{g.pardi, j.hoyer, p.holtz, y.kammerer\}@iwm-tuebingen.de}
	\and
	L3S Research Center, Leibniz University Hannover, Hannover, Germany\\
	\email{rokicki@l3s.de}
}

\maketitle

\begin{abstract}
In informal learning scenarios the popularity of multimedia content, such as video tutorials or lectures, has significantly increased. Yet, the users' interactions, navigation behavior, and consequently learning outcome, have not been researched extensively. Related work in this field, also called \textit{search as learning}, has focused on behavioral or text resource features to predict learning outcome and knowledge gain. In this paper, we investigate whether we can exploit features representing multimedia resource consumption to predict of knowledge gain (KG) during Web search from in-session data, that is without prior knowledge about the learner. For this purpose, we suggest a set of multimedia features related to image and video consumption. Our feature extraction is evaluated in a lab study with 113 participants where we collected data for a given search as learning task on the formation of thunderstorms and lightning.  
We automatically analyze the monitored log data and utilize state-of-the-art computer vision methods to extract features about the seen multimedia resources. Experimental results demonstrate that multimedia features can improve KG prediction. Finally, we provide an analysis on feature importance (text and multimedia) for KG prediction.

\end{abstract}

\section{Introduction} 
\label{sec:introduction}

Traditional information retrieval systems address, in broad terms, the notions of an information need and the corresponding topical relevance of a document~\cite{schutze2008introduction}. This is, however, a simplification: Web search is often used for complex tasks such as learning a new skill or exploring a new topic, i.e., going beyond simple lookup searches.
Broder et al.~\cite{Broder02} distinguish different search \textit{intents}: \textit{transactional} search sessions (e.g., buying something), \textit{informational} sessions for knowledge acquisition, and \textit{navigational} sessions aiming to find a dedicated website.

The research field \textit{search as learning} (SAL) focuses on Web searches with an \textit{informational} intent and explores how they can be supported by information retrieval (IR) systems \cite{kct2017,RiehCHL16}. This entails, for example, the detection of a user's learning intent, the prediction of knowledge state and knowledge gain during search, as well as the adaption of search results according to learning the goals. Thereby, search as learning goes clearly beyond relevance scoring of documents.

Previous work has studied the relationship between learning progress and text content or behavioral features collected from search sessions.
For instance, Collins-Thompson et al.~\cite{collins2016assessing}, studied the influence of distinct query types on knowledge gain, and found that intrinsically diverse queries are correlated with knowledge gain. On the other hand, Syed and Collins-Thompson~\cite{syed2018exploring} explored a range of text and resource-based features and their impact on short-term and long-term learning outcome, but did not investigate multimedia content. Moraes et al.'s~\cite{MoraesPH18} work compared the learning outcome of instructor-designed learning videos against three instances of search ("single-user", "search as support tool", "collaborative search") in order to find the most efficient approach for their learning scenario.
Other work investigated the learning outcomes associated with the consumption of multimedia resources~\cite{shi2019investigating,pardi2020role}. Pardi et al.~\cite{pardi2020role} found that the time users spent on text-dominated websites associates with better learning outcomes compared to videos.
Vakkari~\cite{vakkari2016} provided a structured survey of features indicating learning needs as well as user knowledge and knowledge gain throughout the search process. 
Gadiraju et al.~\cite{gadiraju2018chiir} described the use of knowledge tests to calibrate the knowledge of users before and after their search sessions, quantifying their knowledge gain, and investigated the impact of search intent and search behavior on the knowledge gain of users.
In follow-up work, Yu et al.~\cite{yu2018predicting} utilized interaction features to predict users' knowledge gain in search sessions using supervised machine learning. 
Bhattacharya et al.~\cite{bhattacharya2019measuring} investigated the relationship eye gaze behavior and learning performance in user search. 

In this paper, we investigate the impact of multimedia features on users' knowledge gain in a SAL scenario. 
We conducted a user study that recorded the pre- and post-knowledge states of the participants through multiple-choice questionnaires. 
After the search session, we analyzed all visited Web pages to gather a set of features regarding consumed multimedia content, e.g., document layout, image size and type. 
This novel feature set allows us to investigate the role of multimedia features for knowledge prediction in this SAL scenario. Therefore, we train a supervised learning model (random forest) to predict knowledge gain based on text and multimedia features. Experimental results demonstrate the feasibility of the approach and a feature importance analysis shows that features related to image and video content slightly improve knowledge gain prediction compared to textual- and behavioral feature based methods.

The remainder of the paper is structured as follows: 
Section~\ref{sec:user_study} introduces the setup of our user study, while section~\ref{sec:data_extraction} presents our methodology for multimedia feature extraction. Section~\ref{sec:experiments} reports the results for knowledge gain prediction. Section~\ref{sec:conclusion} concludes the paper and outlines areas for future research.

\section{User Study and Data Collection} 
\label{sec:user_study}

The participants (N=113, $22.86 \pm 2.92$ years old, 96 females) of our lab study were asked to solve a realistic learning task, that is to understand the principles of thunderstorms and lightning. 
The topic of the formation of thunderstorms and lightning has been used in many studies that investigated learning with multimedia (e.g., ~\cite{mayer1998split,schmidt2011role}).
This topic is related to natural sciences and has been chosen since it requires learners to gain knowledge about different physical and meteorological concepts and their interplay. The learning task itself can be classified as a causal task \cite{vangenuchten2012} in which learners need to learn about the causal chains of events. They need therefore to acquire declarative as well as procedural knowledge \cite{anderson2001taxonomy} about different concepts to gain comprehensive knowledge. We believe that this task is a suitable representative for a class of various and similar tasks. For example, comparable causal tasks would be learning about the greenhouse effect or photosynthesis. The acquisition of information about causal tasks can be accomplished through studying different representation formats like text, pictures, videos, or their combined presentation on Web pages. 

\subsubsection*{Technical Setup} All search and learning activities of participants were conducted within a tracking framework that consists of two layers. The SMI (SensoMotoric Instruments) ExperimentCenter (3.7) software enabled us to track participants' activities during Web search through screen recordings and navigation log files. We implemented a second tracking layer in the browser using plugins. These plugins saved all visited HTML files and tracked additionally navigation and interaction data (e.g., mouse movements) in local log files. 
The local and external tracking was realized through JavaScript code integrated into the plugin ``Greasemonkey'' (3.11) running in the browser. To track HTML files, we used the plugin ``ScrapBook'' (1.5.14) which allowed us to automatically and simultaneously save all visited HTML pages (HTML files and folders) seen by the participants. 

\subsubsection*{Knowledge Test} Based on previous work \cite{schmidt2011role} we developed a 10-item multiple-choice knowledge test on the formation of thunderstorm and lightning. To measure participants' pre-knowledge state (pre-KS), this test had to be completed two days before the Web search task, and a second time after the Web search task to measure the post-knowledge state (post-KS). Pre- and post-knowledge states were represented by a score that counted the number of correct answers (out of 10). Knowledge gain (KG) is the difference between the two scores.

\section{Multimedia Feature Extraction} 
\label{sec:data_extraction}
In this section, we outline how we generated the multimedia features based on text, but mainly for image and video content that serve as input for knowledge gain prediction. 
The output of the data logging per user is the input for our feature extraction process. The data logging output consists of a screen recording (MPEG-4 video format (*.mp4)), a timeline of visited Web pages, as well as HTML and CSS files of every visited Web page. In a first step, Web pages are segmented into regions of headlines, (normal) text, images, etc. using a state-of-art method for document layout analysis (Section~\ref{sec:document-layout-analysis}). The image regions are further processed through image type classification to infer the kind of seen content (Section~\ref{sec:image-classification}). In Section~\ref{sec:textual_features} we describe the feature extraction process for text content. Both feature types are then utilized to predict knowledge gain using a random forest classifier. 

\begin{figure*}[!htbp]
\centering
	\includegraphics[width=0.8\textwidth]{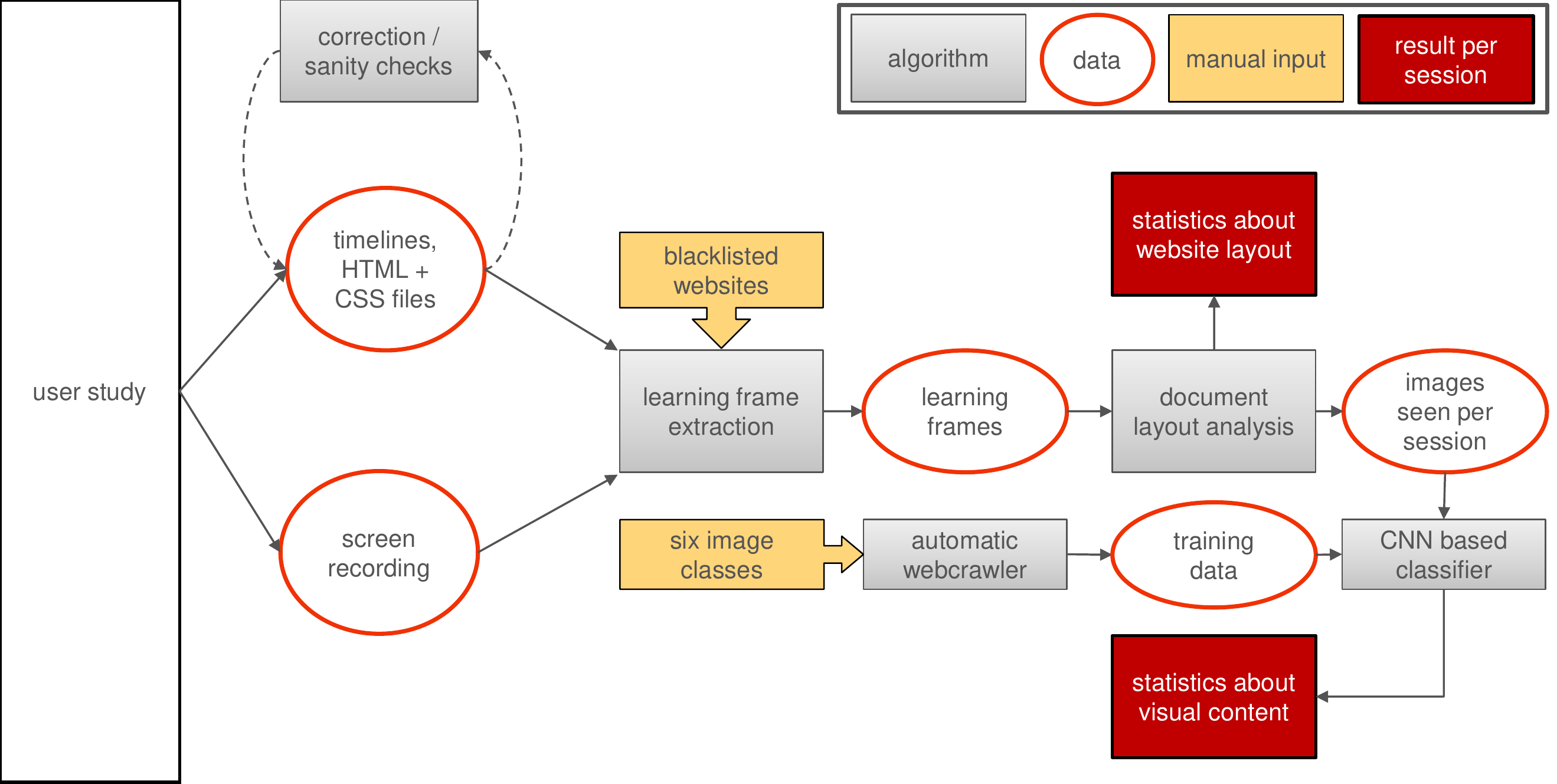}
	\caption{Our multimedia feature extraction pipeline: The only manual inputs are the list of blacklisted websites and the set of image classes. The multimedia features per user (red boxes) as well as textual information described in Section~\ref{sec:textual_features} are the input for our knowledge gain predictor (Section \ref{sec:experiments}).}
	\label{fig:framework}
\end{figure*}

To reconstruct the visited Web pages, we exploit the screen recordings and segment them according to the timeline of the search session. The timeline should reflect the order of Web pages getting into focus, rather than the points in time URLs were opened in their respective browser tabs. In this way, we circumvent the problem of participants opening multiple links from the search result page at once in new tabs, leading to a flawed timeline. An overview of the framework is displayed in Figure~\ref{fig:framework}. 
As shown in Figure~\ref{fig:framework}, the first step separates the total number of  $F$ video frames into $L$ learning relevant and $N$ navigation-related frames, with $N + L = F$. We extract a frame every second of the video ($|F| = 173\,787$), but only keep those where the participant spent time on Web pages related to learning (and not navigating the browser or procrastinating). Thus, we excluded (study-specific) URLs containing \textit{Google}, \textit{TripAdvisor} and \textit{adblock}. This procedure resulted in a total of $119\,164$ (average: 1268 frames per session) learning relevant frames which have to be segmented into pictorial, textual, and background information as described in the next section. 

\subsection{Document Layout Analysise} 
\label{sec:document-layout-analysis}
The goal of this step is to derive features on document layout by automatically dividing each frame $l \in L$ into coherent regions that represent the structure of the page. 
Additionally, the regions should be classified according to their content, e.g., image, text, menu, etc. This procedure is crucial for the image content analysis later but also challenging since the layout and design of the Web pages vary heavily. To address this challenge, we utilize the Mask R-CNN~\cite{matterport_maskrcnn_2017} network architecture, originally implemented for instance (object) segmentation, and fine-tune the provided pre-trained weights of the network. 
We annotated $300$ randomly chosen frames from our user study using the browser-based ''VGG Image Annotator``. Six region classes are distinguished:

\begin{enumerate}
    \item Heading: any headlines or titles that divide the page into sections;
    \item Menu bar: buttons or lists of buttons displayed for navigational purposes;
    \item Content list: enumerations like table of contents or bullet point lists; 
    \item Text: any coherent text block that is not part of headlines or button labels;
    \item Images/Frames: All types of images (no size constraints) from small thumbnails to fullscreen video frames;
    \item Background: everything that does not fit into the five other classes.
\end{enumerate}

\begin{figure}
\centering
\begin{subfigure}[t]{0.48\textwidth}
  \centering
  \includegraphics[width=\linewidth]{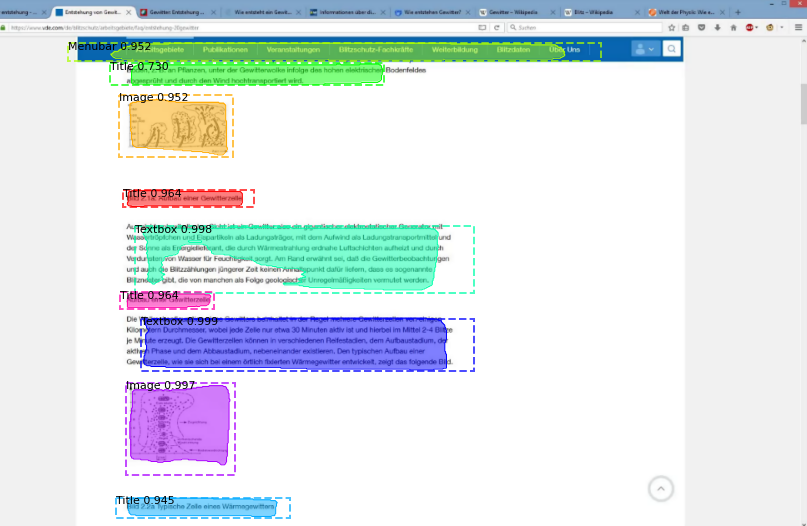}
  \caption{Good example.}
  \label{fig:dla_good_sample}
\end{subfigure}%
\hfill
\begin{subfigure}[t]{0.5\textwidth}
  \centering
  \includegraphics[width=\linewidth]{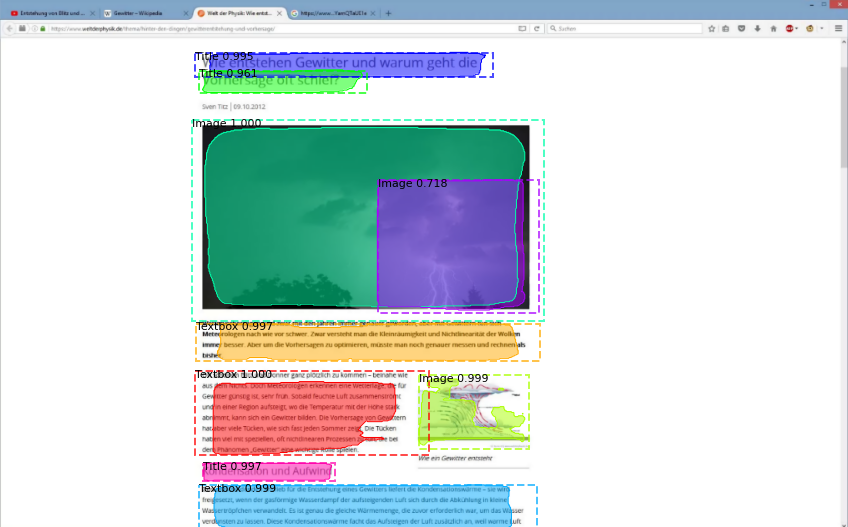}
  \caption{Example with image-in-image effect.}
  \label{fig:dla_image_in_image}
\end{subfigure}
\caption{Two example outputs of the Document Layout Analysis.}
\end{figure}

These classes are supposed to reflect the core parts of a Web page. The JSON (JavaScript Object Notation) style output of the manual annotations was then split into 90\% training and 10\% test data, which we used to fine-tune the fully-connected layers after the pre-trained bounding box detector (i.e., network heads) for 30 epochs with a learning rate of $lr = 0.001$. 
This option is predefined (by the Mask R-CNN authors) by creating the model with parameter $layers =$ ''heads`` and subsequently only retrains the region proposal network (RPN), the classifier, and the mask heads. The resulting network is able to segment our screen recording frames appropriately. An example output is depicted in Figure~\ref{fig:dla_good_sample}.

In addition to our six classes we also compute the average image size per frame since we want to differentiate if a Web page with 20\% visual content contained five small images or a single large one. Another merit of this feature is its ability to also indirectly measure the viewing time of videos since it is difficult to measure this feature directly with satisfactory accuracy. For instance, embedded videos on Web pages other than YouTube cannot always be captured. 
With this, our document layout features per frame $i$ are represented as a vector $d_i$ containing six percentages and a scalar. 
\begin{equation}
d_i = (head_i, menu_i, conlist_i, text_i, img_i, bg_i, \overline{imgsize}_i), \forall i \in L. 
    \label{formula:dla_per_frame}
\end{equation}
Lastly, the results per frame in equation~\ref{formula:dla_per_frame} are summed up for all seen learning frames $i \in L$ and divided by $|L|$ yielding seven features per participant $p$.
\begin{equation}
    d_p =  (\sum_{i=0}^{|L|} \frac{head_i}{|L|}, \sum_{i=0}^{|L|} \frac{menu_i}{|L|}, ... , \sum_{i=0}^{|L|} \frac{\overline{imgsize}_i}{|L|})
\label{formula:dla_per_participant}
\end{equation}
We identified a total of $755\,756$ bounding boxes that belonged to the ''Images/Frames`` class, which has around five samples per frame on average. This appears to be a lot at first, but has a simple explanation. Every (Web page) frame that is recorded when watching a (non-maximized) YouTube video contains 10 thumbnails of other recommended videos. In order to not skew the results heavily towards this large number of irrelevant images, we filtered them out if their height or width is below 100 pixels (full image resolution was 1280x800). The remaining samples will be further examined regarding their shown content.

\subsection{Image Type Classification}
\label{sec:image-classification}
This section briefly outlines how the images detected in the document layout analysis are examined regarding their content. We aim to predict the given \textbf{type} of an image. 
To the best of our knowledge, there is no comprehensive and task-specific taxonomy of image types that can be directly applied to the learning task of our study.
Therefore, we focused on covering all topic-relevant categories to learn which type of images a learner saw when searching for the formation of thunderstorms. As a result, our set of image type classes consists of Infographics, Indoor Photo, Maps, Outdoor Photo, Technical Drawings, and Information Visualization.  The class \textit{Information Visualization} has a specific role. Images that are composites or hybrids of common visualization types are hard to assign to a unique class. 
For this reason, we merge all forms of \textit{Information Visualizations} into one class and use it as a fallback class to gather all frames that are otherwise hard to assign.

The implementation was done in Keras, using a MobileNet~\cite{howard2017mobilenet} architecture with default parameters. We utilized a Google image crawler to gather $18\,773$ unique training samples which we split into 90\% training and 10\% test data. Three volunteers manually labeled the test data and achieved an intercoder agreement of $\alpha = 0.85$ (across all annotators, samples, and classes) according to Krippendorff's alpha~\cite{krippendorff1970estimating}.
Finally, the classifier achieved an accuracy of $87.15\%$ on this human-verified test set. The accuracy is sufficient for our task of knowledge gain prediction, as it is confirmed by the experimental results in Section~\ref{sec:experiments}.

The features do not represent the number of images seen per class, because a consequence of our frame-wise extraction is that the same image gets extracted multiple times. Instead we analyse the image in every frame again and report the fraction of the image types as a percentage. The idea is to weight the content according to the duration the images have been seen by the learner. The feature vector $v_p$ representing the image types seen by participant $p$ is defined as follows:

\begin{equ}[!ht]
\begin{equation}
v_p = (\sum_{l=0}^{|L|} \sum_{n=0}^{N_l} p(Info.-Vis.), ...\,, \sum_{l=0}^{|L|} \sum_{n=0}^{N_l} p(Techn.\,Draw.))
\end{equation}
\captionsetup{labelformat=empty, format=plain}
\caption{Feature vector for the six image types seen per participant. $p(<class>)$ is the pseudo-probability given by the softmax layer. $N_l$ is the number of images detected in frame $l$.}
\end{equ}

\subsection{Text Features}
\label{sec:textual_features}
In total, we used a set of 110 features to represent textual information\footnote{Full feature list at: \href{https://www.dropbox.com/s/l8zy4kn79c1ytc3/textual_information_based_features.pdf?dl=0}{Dropbox-Link}}, taking into account document complexity, HTML structure, and linguistic aspects. 

\textbf{Document Complexity Features.}
Based on the assumption that the document complexity is correlated with the user's knowledge state on a topic, we have extracted several features related to document complexity. Motivated by previous work~\cite{eickhoff2014lessons} and our investigation of the data, we extracted the number of words ($c\_word$), length of words ($c\_char$), and length of sentences ($c\_sentence$) as features. Related work~\cite{heilman2007combining} suggests that the syntactic structure of a document, which can be represented by the ratio of the number of nouns, verbs, adjectives, or \textit{other words} 
to the total number of words ($c\_\{noun, verb, adj, oth\}$) is likely to imply the complexity of its content. 

There are several widely used metrics for assessing the readability or complexity of a textual document, which have been studied to be correlated with user's knowledge level~\cite{horne2017just}. We used Gunning Fog Grade\footnote{\url{http://gunning-fog-index.com/}}($c\_gi$), SMOG~\cite{mc1969smog} ($c\_smog$) and Flesch-Kincaid Grade~\cite{kincaid1975derivation} ($c\_fk$) as features. 

\textbf{HTML Structural Features.} 
A possible explanation of the finding, that there is a negative association between the number of hyperlinks embedded in a Web page and users' KG~\cite{destefano2007cognitive}, is that people may not focus on the content in the presence of too many embedded links. Hence, we extract the feature $h\_link$ by quantifying the number of outbound links (i.e., the \textit{$<a>$} elements in our case). Furthermore, we extract features that might indicate the readability of a Web page based on HTML tags, namely, the average length of each paragraph ($h\_p$), the \textit{$<ul>$} elements embedded ($h\_oth\_ul$), and the number of scripts ($h\_script$).

\textbf{Linguistic Features.}
Related work~\cite{horne2017just} suggests that the number of words on Web pages that are correlated with different psychological processes and basic sentiment can influence a learner's cognitive state. The writing style could also affect the readability of a learning resource and the engagement of readers. 
Motivated by the above observations, we used the 2015 Linguistic Inquiry and Word Count (LIWC) dictionaries\footnote{\url{http://liwc.wpengine.com/compare-dictionaries/}} to compute linguistic features that reflect the psychological processes, sentiment, and the writing style of Web page content. The features of this type are prefixed with $l\_$ in the remainder of the paper.

\section{Experimental Results for Knowledge Gain Predcition}
\label{sec:experiments}
In this section, we report experimental results for the task of knowledge gain prediction utilizing the features from Section \ref{sec:data_extraction}. 
Our experimental dataset consists of 113 search sessions. On average, users have issued 11.1 queries and browsed 25.4 Web pages in each session. 
There was a significant increase in learners' knowledge on average (KG = $2.15 \pm 1.84$ for a full score of 10) after the search phase. The effect size for knowledge gain was large (Cohen's d = 1.29). The average pre-knowledge score was $5.22 \pm 1.76$ and post-knowledge was $7.37 \pm 1.6$. 

\subsection{Experimental Setup}
The goal of our study is to predict knowledge gain in informal search sessions and to investigate the impact of text and multimedia resource features. We model KG prediction as a classification task and use random forest as a supervised learning approach.
We aim for a fair comparison with the state of the art in users' knowledge gain prediction in Web search. Thus, we follow the same experimental setup as used by Yu et al.~\cite{yu2018predicting}, in particular for the assignment of labels, the applied classifier, and its parameter tuning, unless other settings are denoted. 

\textbf{Ground Truth Data:} We group a search session into one of three KG classes according to the measured knowledge gain $X$ based on the \textit{Standard Deviation ($\sigma$) Classification} approach. The classes are defined as follows: 1.) \textit{Low} KG, if $X < \overline{X} - \frac{\sigma}{2}$; 2.) \textit{Moderate} KG, if $\overline{X} -\frac{\sigma}{2} < X < \overline{X} + \frac{\sigma}{2}$; and 
3.) \textit{High} KG, if $X > \overline{X} + \frac{\sigma}{2}$. According to this approach, our dataset consists of 44 low, 42 moderate, and 27 high knowledge gain sessions. 

\textbf{Classifier:} Random forest has shown to be the most effective classifier for knowledge gain prediction~\cite{yu2018predicting} and it allows for the analysis of feature importance.
Hence, we adopt a random forest classifier and tune the hyperparameters for accuracy using grid search. 
For our experiments, we used the \textit{scikit-learn} library for Python ({\url{https://scikit-learn.org/}}).

\textbf{Metrics:} After tuning the hyper-parameters of each classifier, we run 10 repetitions of 10-fold cross-validation (90\% train, 10\% test) and evaluate the classification results of each classifier according to the following metrics:

\begin{itemize}[leftmargin=*, nosep]
\item \textbf{Accuracy ($Accu$) across all classes:} percentage of search sessions that were classified with the correct class label.
\item \textbf{Precision ($P$), Recall ($R$), F1 ($F1$) score of class $i$:} the standard precision, recall and F1 score on the prediction result of each class $i$.
\item \textbf{Macro average of precision ($P$), recall ($R$), and F1 ($F1$):} the average of the corresponding scores across three classes.
\end{itemize}

\subsubsection{Classification Results. }\label{sec:classification_result}

The performance of the random forest classifier using textual features (TI), multimedia features (MI), as well as their combination is shown in Table \ref{tab:classification_result}. We also present the performance of the random forest classifier using behavioral features (the approach used by \cite{yu2018predicting}) on our ground truth dataset in Table~\ref{tab:classification_result} for reference. 
The results for all four test feature types are in a comparable range. When using only textual features (TI) or multimedia features (MI), results are slightly worse than the state-of-the-art approach~\cite{yu2018predicting}, which uses behavioral features.
Our approach utilizing features from both categories (MI\&TI) has achieved the best performance concerning overall accuracy, indicating that the combination of textual and multimedia features has the potential to improve knowledge gain prediction. 
Comparing the performance for the different classes, the classifier performs better on low and moderate knowledge gain classes when using features from both categories. A potential reason for this result is that the high knowledge gain class has the least amount of training data in our ground truth dataset.

Please note that we have achieved comparable performance to the state of the art~\cite{yu2018predicting} with less training data (113 sessions versus 468 sessions) and more unbalanced classes. 
As shown in Table \ref{tab:classification_result} the combination of multimedia and textual information (MI\&TI) is able to outperform using behavior features at 85\% confidence level in terms of accuracy. However, please note that we did not extract all of the behavioral features introduced in the related work, in particular, the features relevant to the clicking on the search results page were not recorded in our study.
Since we focus on understanding the influence of textual and multimedia resource content on users' knowledge gain during the search, the classification model and the analysis of user behavior features are out of the scope of this work. We list the results of the classifier trained on user behavior features as evidence that our classification has reached satisfying performance. 

\begin{table*}[!htbp]
	\small
	\centering
 	\scalebox{0.95}{
	\begin{tabular}{|p{1.4cm}|p{0.75cm}p{0.75cm}p{0.7cm}|p{0.75cm}p{0.75cm}p{0.7cm}|p{0.75cm}p{0.75cm}p{0.7cm}|p{0.75cm}p{0.75cm}p{0.7cm}|p{0.8cm}|}
	\hline
  & \multicolumn{3}{c|}{\textbf{Low}}& \multicolumn{3}{c|}{\textbf{Moderate}}& \multicolumn{3}{c|}{\textbf{High}}& \multicolumn{3}{c|}{\textbf{Macro average}} & \textbf{All} \\
   \textbf{Features}& \textbf{P}& \textbf{R}& \textbf{F1}& \textbf{P}& \textbf{R}& \textbf{F1}& \textbf{P}& \textbf{R}& \textbf{F1}& \textbf{P}& \textbf{R}& \textbf{F1} &\textbf{Accu} \\
   \hline
MI\&TI   & \textbf{41.5} & \textbf{52.0} & \textbf{46.1} & \textbf{39.1} & \textbf{40.0} & \textbf{39.5} & 28.4          & 14.8          & 19.1          & 36.4          & 35.6          & 34.9          & \textbf{38.7} \\\hline
TI       & 39.9          & \textbf{52.0} & 45.0          & 36.6          & 33.8          & 35.0          & 28.9          & 17.4          & 21.5          & 35.1          & 34.4          & 33.8          & 37.0          \\\hline
MI       & 38.0          & 38.0          & 37.9          & 38.0          & 38.1          & 38.0          & 30.8          & \textbf{31.1} & \textbf{30.8} & 35.6          & 35.7          & 35.6          & 36.4          \\\hline
\cite{yu2018predicting} & 39.7          & 47.0          & 43.0          & 37.4          & 39.5          & 38.4          & \textbf{34.9} & 21.1          & 26.0          & \textbf{37.3} & \textbf{35.9} & \textbf{35.8} & 38.1       \\
   \hline
	\end{tabular}}
\caption{Results of knowledge gain prediction (in \%) using text (TI) and multimedia features (MI), and comparing them with the state of the art~\cite{yu2018predicting}. }
\label{tab:classification_result}
\end{table*}

\subsubsection{Feature Importance}
\label{sec:feature_importance}
To analyze the usefulness of individual features, we make use of the \textit{Mean Decrease in Impurity (MDI)} metric based on the random forest model. The metric MDI is defined as the total decrease in node impurity (weighted by the probability of reaching that node) averaged over all trees of the ensemble \cite{breiman2017classification}. 
Due to the space limitation, we only list and discuss the 20 features (Table \ref{tab:feature_importance}) having the highest and lowest MDI values in the paper.

We observe that six out of 10 features with the highest importance are textual features. This is to be expected because, first, there are more textual features (110) than multimedia features (13), and, second, with recent advances in natural language processing techniques, we were able to design more sophisticated textual features such as the complexity of language and sentiment behind words. In contrast, it is still more challenging to analyze the semantics of multimedia data. Nevertheless, results indicate that the 13 multimedia features have shown promising importance for the classification, with \textit{Heading}, $\overline{imgsize}$, \textit{Menu Bar}, \textit{Infographic}, \textit{Technical Drawing} and \textit{Outdoor} rank at 4, 5, 8, 9, 13, 15, respectively, among the 123 features in total. None of the multimedia features falls into the 10 least important features according to MDI. 
Among the six textual features with the highest importance, five are linguistic-based, while the remaining one is related to document complexity  (SMOG Readability). 

\begin{table}[]
\centering
\begin{tabular}{|l|ll|ll|}
\hline
              & \multicolumn{2}{c|}{\textbf{Highest}} & \multicolumn{2}{c|}{\textbf{Lowest}} \\
\textbf{Rank} & \textbf{feature}    & \textbf{MDI}   & \textbf{feature}    & \textbf{MDI}  \\ \hline
\textbf{1}  & l\_home         & 0.039 & l\_affect    & 0.004 \\
\textbf{2}  & l\_relig        & 0.030 & l\_Tone      & 0.004 \\
\textbf{3}  & l\_certain      & 0.018 & l\_power     & 0.004 \\
\textbf{4}  & Heading           & 0.018 & l\_AllPunc   & 0.003 \\
\textbf{5}  & $\overline{imgsize}$      & 0.016 & h\_vid       & 0.003 \\
\textbf{6}  & c\_smog         & 0.015 & l\_filler    & 0.003 \\
\textbf{7}  & l\_focuspresent & 0.015 & l\_sad       & 0.003 \\
\textbf{8}  & Menubar         & 0.015 & h\_aud       & 0.003 \\
\textbf{9}  & Infographic     & 0.014 & l\_Authentic & 0.002 \\
\textbf{10} & l\_netspeak     & 0.014 & h\_obj       & 0.001 \\\hline
\end{tabular}
\caption{Features with highest and lowest MDI importance scores. }\label{tab:feature_importance}
\end{table}

\section{Conclusions}
\label{sec:conclusion}
In this paper, we have investigated whether features describing multimedia resource content can help predict users' knowledge gain in a search as learning task. Our results are based on a lab study with N=113 participants, where we recorded the individuals' behavior and the accessed Web resources. Afterwards, we applied computer vision methods to screen recordings to segment the seen Web pages into meaningful regions, and then further classified the image regions into task-specific image types. Finally, we used the textual and multimedia features to classify the knowledge gain of the participants.

The combination of our different feature categories and detailed feature importance assessment showed that our approach can serve for knowledge gain prediction based on viewed resource content, which potentially can help improve a learning-oriented search result ranking (if content features are used accordingly). Although the classification accuracy is on a moderate level in terms of recall and precision, they suggest that knowledge gain is predictable. Particularly image and video features improved the classification notably when used jointly with text-based features. To the best of our knowledge, that was the first study that analyzed the importance of multimedia features in a SAL scenario. 

Although the number of participants in our study is already higher than in the majority of previous studies in controlled lab settings, our current dataset is limited by the fact that only one learning task has been studied. In the future, we aim to conduct additional studies on diverse learning topics, to receive further insights into the relationship between features of learning resources used and knowledge gain.


\bibliographystyle{splncs04}
\bibliography{short_references}

\end{document}